\documentclass[aps,prd,twocolumn,superscriptaddress,nofootinbib,floatfix]{revtex4-2}

\usepackage{amsmath,amssymb,bm}
\usepackage{graphicx}
\usepackage[colorlinks=true,linkcolor=blue,citecolor=blue,urlcolor=blue]{hyperref}

\newcommand{\GF}{G_F}
\newcommand{\nn}{n_n}
\newcommand{\nnu}{n_\nu}
\newcommand{\Enu}{E_\nu}
\newcommand{\munu}{\mu_\nu}
\newcommand{\rhat}{\hat{\mathbf r}}
\newcommand{\khat}{\hat{\mathbf k}_0}
\newcommand{\bsig}{\bm\sigma}

\begin{document}

\title{Non-accumulation of background-enhanced neutrino forces\\
in proto-neutron stars}

\author{Himanshu Bora}
\email{hb@kamrupcollege.ac.in}
\affiliation{Department of Physics, Bhattadev University, Pathsala, Bajali, 781325, Assam, India}
\affiliation{Department of Physics, Kamrup College, Chamata, Nalbari, 781306, Assam, India}

\author{Anupam Deka}
\email{anupamdeka585@gmail.com}
\affiliation{Department of Physics, Bhattadev University, Pathsala, Bajali, 781325, Assam, India}
\affiliation{Department of Physics, Nabajyoti College, Kalgachia, Barpeta, 781319, Assam, India}

\author{Debajyoti Dutta}
\email{phy.debajyoti@bhattadevuniversity.ac.in}
\affiliation{Department of Physics, Bhattadev University, Pathsala, Bajali, 781325, Assam, India}

\begin{abstract}
In a neutrino background, the vacuum two-neutrino exchange potential $V\propto \GF^2/r^5$ acts as a much longer-ranged interaction, scaling as $V\sim1/r$ for a directional background. A proto-neutron star (PNS) hosts one of the densest astrophysical neutrino environments known, and naturally provides a setting in which to test whether this pairwise enhancement produces a macroscopic Hartree contribution. We show that this does not happen. In the spin-independent sector, averaging over pair orientations projects the background onto its angular monopole. The bulk Hartree contribution then saturates instead of growing with system size. We also find that the spin-dependent parity-conserving and parity-violating terms do not generate macroscopic growth under the PNS mean-field conditions considered here. Thus, although the pair potential can be strongly enhanced, the resulting bulk interaction remains very small. The suppression comes from the angular and oscillatory structure of the kernel, not from screening. Within the Standard Model mechanism considered here, the effect is not important for the PNS equation of state or for events like pulsar kicks.
\end{abstract}

\maketitle

\section{Introduction}
\label{sec:intro}

The Standard Model (SM) predicts a long-range force between fermions generated by the exchange of a neutrino pair~\cite{Feinberg1968,Feynman,Feinberg1989,Hsu1994}. In vacuum, and at distances where the neutrino masses can be neglected, the spin-independent potential is
\begin{equation}
V_0(r)=\frac{\GF^2 g_V^1 g_V^2}{4\pi^3 r^5},
\qquad
m_{\chi}^{-1}\ll r\ll m_\nu^{-1},
\label{eq:vacuum}
\end{equation}
where $\GF$ is the Fermi constant, $g_V^{1,2}$ are the vector weak couplings of the two fermions, $m_\chi$ denotes the characteristic mass of the external fermions, and $m_\nu$ is the neutrino mass. As a loop-induced interaction, it is extremely weak and is far below the gravitational interaction between elementary particles already at nanometre separations.

The situation changes in a neutrino background. Replacing one
propagator by its background-modified counterpart
~\cite{Notzold1988,Landsman1987} introduces an on-shell background
contribution. In the plane-wave treatment of a directional neutrino background, the vacuum two-neutrino exchange potential
$V\propto G_F^2/r^5$ acquires a much longer-ranged
$V\sim1/r$ component and greatly enhance the
pair potential~\cite{Ghosh2023,Ghosh2024}. Finite wave packets and
averaging over angular or spatial extent can substantially reduce the directional enhancement~\cite{Ghosh2023,Blas2023,VanTilburg2024}. In the plane-wave pair-potential estimate, the enhancement can reach
about twenty orders of magnitude for reactor or solar neutrino
backgrounds, and the same mechanism has since been generalized to axion backgrounds~\cite{Grossman2026}, dark relics~\cite{Barbosa2025}, and generic quadratically coupled mediators~\cite{VanTilburg2024}. Neutrino-mediated quantum forces have also been studied in Casimir geometries~\cite{Costantino2020} and in dark-matter applications without an ambient background~\cite{Orlofsky2021,Coy2022}. More recently, Ref.~\cite{Ittisamai2026} studied the modification of dark-matter self-interactions by the cosmic neutrino background.

A proto-neutron star (PNS) hosts one of the densest astrophysical neutrino environments known. During the first $\sim10\,$s after collapse, the core traps neutrinos at $\nnu\sim10^{38}\,\mathrm{cm}^{-3}$ with chemical potential $\munu\sim200\,$MeV~\cite{Prakash1997,Alford2019}, just outside the neutrinosphere, a radially beamed free-streaming neutrino population has number density $\sim10^{32}\,\mathrm{cm}^{-3}$. These conditions make the PNS an especially stringent environment in which to test whether the background-induced enhancement can accumulate macroscopically. The per-pair enhancement is very large. In the free-streaming region outside the neutrinosphere the directional pair potential exceeds the neutron–neutron gravitational potential by $\sim8\times10^{16}$. Therefore, a pairwise estimate alone could suggest a potentially significant macroscopic effect. 

A related issue arose in the 1990s, when multibody neutrino exchange in neutron stars was argued to produce an anomalously large stellar self-energy~\cite{Fischbach1996}. Subsequent analyses resolved that through several complementary approaches: by Pauli principle and thermal blocking of the neutrino sea~\cite{Smirnov1996}, by explicit resummation of the many-body series~\cite{Abada1996}, by direct recalculation of the self-energy~\cite{Kachelriess1998}, by a nonperturbative treatment showing that the perturbative expansion itself fails for large systems while the exact one-loop ground-state energy remains well behaved~\cite{Kiers1998}, and by infrared-safe finite-size analyses~\cite{Abada1998,Arafune1998}. The common conclusion was that the large stellar self-energy is not physical.

In this work, we address the corresponding question for background-enhanced two-neutrino exchange and find no macroscopic accumulation in the PNS regimes considered. Two regimes are relevant to a PNS. In the deep core the trapped neutrino gas is approximately isotropic and strongly degenerate ($\munu\gg T$). Outside the neutrinosphere the flux is free-streaming and directional. We treat both. The non-accumulation follows from the structure of the StandardModel background-induced kernels considered here rather than from screening. The same on-shell contribution that enhances the pair potential also makes the position-space kernel oscillatory. Whether the corresponding bulk integral accumulates then depends on the angular and radial structure of the kernel. The central observation in our paper is that for the Standard Model kernels considered here, that structure is sufficient to prevent macroscopic growth.

Let $K^\mu=(K^0,\mathbf k)$ denote the background-neutrino four-momentum, with $k\equiv|\mathbf k|$. For the spatial momentum transfer we write $\mathbf q=\rho\,\hat{\mathbf q}$, where $\rho\equiv|\mathbf q|$. The enhancement originates in the on-shell delta function $\delta(K^2-m_\nu^2)$ in the background propagator, which produces denominators of the form $\rho^2-4(\mathbf k\cdot\hat{\mathbf q})^2$ with singularities on the real $\rho$ axis. By contrast, the vacuum two-neutrino amplitude is controlled by the two-particle branch cut discussed in Sec.~\ref{sec:statement}. The resulting background contribution is oscillatory, with a characteristic scale set by the background-neutrino momentum. Under the Hartree orientation average, all higher angular multipoles of the spin-independent background distribution are projected out; only its monopole spectrum survives. Previous studies established background-enhanced pair potentials in neutrino and related backgrounds ~\cite{Ghosh2023,VanTilburg2024,Ghosh2024,Barbosa2025,Ittisamai2026}. Finite wave packets and averaging over angular or spatial extent
can substantially reduce the directional enhancement
~\cite{Ghosh2023,Blas2023,VanTilburg2024}. Here we evaluate the bulk Hartree integral $\int d^3r\,V$ over pair orientations and spin sectors in a PNS, together with its scaling in $R$ as this quantity is relevant to bulk observables.

The remainder of the paper establishes this quantitatively. We first derive the bulk Hartree result and then evaluate it for the trapped-core and free-streaming PNS regimes. Section~\ref{sec:setup} fixes conventions. Section~\ref{sec:theorem} derives the non-accumulation results, Secs.~\ref{sec:numbers} and~\ref{sec:robustness} quantify the PNS benchmarks and assess their robustness, and Secs.~\ref{sec:hierarchies},~\ref{sec:kicks},~\ref{sec:bsm},~\ref{sec:lab} discuss competing weak effects, pulsar kicks, BSM extensions, and laboratory implications.

\section{Setup}
\label{sec:setup}

We use the four-fermion interaction
\begin{equation}
\mathcal L
=
\frac{\GF}{\sqrt2}
\left[\bar\nu\gamma^\mu(1-\gamma_5)\nu\right]
\left[
\bar\chi\gamma_\mu
\left(g_V^\chi+g_A^\chi\gamma_5\right)\chi
\right],
\end{equation}
where $\chi$ denotes a generic external fermion. For a pair of external fermions we denote their weak couplings by $g_{V,A}^{1,2}$. For neutrons, $g_V^n=-1/2$ for all neutrino flavors, so that $g_V^1g_V^2=1/4$ for a neutron pair. With the $+g_A^\chi\gamma_5$ convention used above, the axial neutron coupling is $g_A^n=+g_A^{(N)}/2\simeq+0.635$, where $g_A^{(N)}\simeq1.27$. The neutral-current scattering coefficients used in Sec.~\ref{sec:hierarchies} are therefore $c_V=g_V^n=-1/2$ and $c_A=-g_A^{(N)}/2\simeq-0.635$.

We work in natural units. We quote number densities in $\mathrm{GeV}^3$ ($1\,\mathrm{GeV}^3=1.301\times10^{41}\,\mathrm{cm}^{-3}$) and energy densities or pressures in $\mathrm{GeV}^4$ ($1\,\mathrm{GeV}^4 =2.085\times10^{38}\,\mathrm{dyn\,cm^{-2}}$).

For a homogeneous neutron medium of number density $n_n$, let $V(\mathbf r)$ denote the neutrino-mediated pair potential between two neutrons separated by $\mathbf r$. The corresponding Hartree interaction energy density and one-body mean-field potential are
\begin{equation}
\mathcal E_{2\nu}
=
\frac12 n_n^2\!\int d^3r\,V(\mathbf r),
\qquad
U_{2\nu}
=
n_n\!\int d^3r\,V(\mathbf r).
\label{eq:hartree}
\end{equation}
For an unscreened long-range $1/r$ pair potential, such as the Newtonian gravitational potential, $\int d^3r\,V\propto R^2$. The Hartree contribution per particle therefore grows quadratically with the system size. The background-enhanced neutrino interaction considered here does not exhibit this $R^2$ growth.

For the PNS benchmarks considered below, the characteristic products $k_{\rm bkg}R$ are of order $10^{18}$--$10^{19}$, where $k_{\rm bkg}$ denotes the relevant background-neutrino momentum scale. The system is therefore deep in the short-wavelength regime defined in Eq.~\eqref{eq:validity}. This condition alone is not sufficient for non-accumulation; the required kernel-specific cancellations are derived in Sec.~\ref{sec:theorem}.

\section{Bulk non-accumulation of the Hartree Term}
\label{sec:theorem}

\subsection{Definition and domain of validity}
\label{sec:statement}

We define non-accumulation and specify the regime in which the result applies before evaluating the Standard Model kernels.

In vacuum, the two-neutrino exchange amplitude has a two-particle
branch cut beginning at the neutrino-pair threshold
$t=(m_i+m_j)^2$. For nonzero neutrino masses, the corresponding
nonanalyticity gives an exponentially suppressed large-distance
contribution, $\propto e^{-(m_i+m_j)r}$; in the massless limit the
threshold moves to the origin and the familiar $1/r^5$ behavior is
recovered. In a neutrino background, by contrast, the on-shell part of
the modified propagator, $\propto\delta(K^2-m_\nu^2)$, produces
denominators of the form
\begin{equation}
\rho^2-4(\mathbf k\cdot\hat{\mathbf q})^2,
\end{equation}
with singularities on the real $\rho$ axis. The corresponding
position-space contribution is therefore oscillatory rather than
exponentially suppressed. Its characteristic wavenumber is set by the
background-neutrino momentum. This same on-shell contribution is
responsible for the enhancement of the background-induced potential
from the vacuum $1/r^5$ behavior to the directional $1/r$ behavior.

The consequence for a bulk observable follows from
Eq.~\eqref{eq:hartree}. A non-oscillating $1/r$ kernel gives
$\int d^3r\,V\propto R^2$. Oscillation by itself does not guarantee
convergence of the bulk integral. As a local homogeneous Hartree
counterexample, consider the isotropically oscillating kernel
$V(r)=\cos(kr)/r$, which has the same qualitative
real-pole/oscillatory behavior. Then
\begin{equation}
\int_{|\mathbf r|<R}\!\!\! d^3r\,\frac{\cos kr}{r}
=
4\pi
\left[
\frac{R\sin kR}{k}
+
\frac{\cos kR-1}{k^2}
\right],
\label{eq:counterexample}
\end{equation}
whose envelope grows as $O(R/k)$ however large $kR$ becomes.
Oscillation alone therefore reduces the $R^2$ growth to $R$, but does
not remove it.

Eliminating the residual $R$-dependence requires additional suppression
from the kernel structure. For the Standard Model spin-independent
kernel, the directional and isotropic cases are not independent at the
Hartree level: the orientation average maps the former onto the latter.

For the spin-independent sector one can obtain an exact finite-$R$
expression. Equation~\eqref{eq:avgidentity} below shows that averaging
over pair orientations projects the background onto its angular
monopole, so a directional beam is bulk-equivalent to an isotropic shell
with the same spectrum. All non-monopole angular information is removed
by this orientation average. The remaining contribution is governed by
the universal form factor $\mathcal F(kR)$ of
Eq.~\eqref{eq:formfactor}: its small-$kR$ limit reproduces ordinary
$R^2$ accumulation, while its large-$kR$ limit saturates. The
counterexample in Eq.~\eqref{eq:counterexample} survives because its
phase carries no corresponding angular structure and therefore receives
no such suppression.

We call an interaction \emph{non-accumulating} when its Hartree
integral $\int d^3r\,V$ approaches a finite, $R$-independent limit as
$R\to\infty$. For the Standard Model SI kernel, the finite-$R$
Hartree integral can be obtained exactly. Equation~\eqref{eq:master} is
exact at finite $R$ for the local homogeneous Hartree construction
defined below, and its $R\to\infty$ limit,
Eq.~\eqref{eq:SIlimit}, is finite and requires no infrared regulator.

For a single spectral mode, the short-wavelength condition is
\begin{equation}
kR\gg1,
\label{eq:validity}
\end{equation}
meaning that its oscillation is well resolved across the approximately
homogeneous region of size $R$. For a narrow background spectrum we
denote the characteristic momentum by $k_{\rm bkg}$ and write this as
$k_{\rm bkg}R\gg1$. For a broad spectrum there is no unique
$k_{\rm bkg}$; the condition must then be understood mode by mode.

Spectral weight satisfying $kR\ll1$ instead obeys
\begin{equation}
\mathcal F(kR)
\simeq
\frac{2}{3}(kR)^2
\end{equation}
by Eq.~\eqref{eq:formfactor}, and therefore gives ordinary $R^2$
accumulation. The short-wavelength condition is not sufficient by
itself: the counterexample in Eq.~\eqref{eq:counterexample} also
satisfies $kR\gg1$ but still grows with $R$.

The parameter $kR$ therefore determines whether a given spectral mode
lies in the long- or short-wavelength regime, while actual saturation
also depends on the angular and radial structure of the kernel. In both
PNS regimes considered below, the short-wavelength condition is
satisfied by an enormous margin. In the free-streaming region,
\begin{equation}
k_{\rm bkg}R
=
1.8\times10^{18},
\qquad
E_\nu=12\,\mathrm{MeV},
\qquad
R=R_\nu,
\end{equation}
whereas in the trapped core
\begin{equation}
k_{\rm bkg}R
=
3.0\times10^{19},
\qquad
\mu_\nu=200\,\mathrm{MeV}.
\end{equation}

By contrast, for a cold non-relativistic background the inverse
wavenumber can be macroscopic. For axion dark matter with
$m_a=10^{-5}\,\mathrm{eV}$ and $v\sim10^{-3}$,
\begin{equation}
k_{\rm bkg}\simeq m_av,
\qquad
\frac{1}{k_{\rm bkg}}
\simeq
20\,\mathrm{m},
\end{equation}
so that $k_{\rm bkg}R\simeq0.05$ in a metre-scale apparatus and the
short-wavelength analysis does not apply. For comparison, the relic
neutrino background lies much closer to the crossover, with
$1/k_{\rm bkg}\simeq0.37\,\mathrm{mm}$, consistent with the
millimetre scale discussed in Ref.~\cite{VanTilburg2024}.

The analysis below establishes non-accumulation for all three Standard
Model spin sectors in the isotropic regime. For a directional
background, the result is established for the homogeneous Hartree
configurations relevant to the PNS treatment; the leading
parity-violating argument additionally uses collinear mean-field
polarization, as discussed in Sec.~\ref{sec:spin}.

We do not claim non-accumulation for arbitrary background-induced
forces. An extension inherits the result only if the relevant modes lie
in the short-wavelength regime and the angular or radial structure of
the kernel removes the residual $R$-dependence.
Section~\ref{sec:bsm} discusses this point further.

\subsection{Sum rules}
\label{sec:sumrules}

The kernels governing the different spin structures are not
independent. Several of them satisfy exact algebraic relations that
simplify the subsequent analysis.

For a directional background, define
\begin{equation}
\xi\equiv
\hat{\mathbf k}_0\cdot\hat{\mathbf q},
\end{equation}
where $\hat{\mathbf k}_0$ is the background-neutrino direction and
$\hat{\mathbf q}$ is the direction of the spatial momentum transfer.
Comparing the angular prefactors
$\tfrac{3+\cos 2\theta}{2}$ and
$\tfrac{5-\cos 2\theta}{2}$ and the weights
$(1\mp\xi^2)$ in the defining integrals of
Ref.~\cite{Ghosh2024} gives the exact identity
\begin{equation}
\mathcal I_a
=
2\,\mathcal I^{\rm PV}
-
\mathcal I^{\rm SI}.
\label{eq:sumrule1}
\end{equation}
Here the superscript ${\rm PV}$ denotes the parity-violating kernel;
later, $\operatorname{PV}$ preceding an integral will denote a
Cauchy principal value.

For an isotropic background, the corresponding relation among the
$\mathcal J$-factors is
\begin{equation}
\mathcal J_a
=
\mathcal J_b+\mathcal J_c .
\label{eq:sumrule2}
\end{equation}
This identity holds pointwise in the background momentum $k$, before
any spectrum is specified. Writing
\begin{equation}
\mathcal J_i(r)
=
r\int_0^\infty dk\,f_0(k)\,\iota_i(k,r),
\end{equation}
the massless integrands are
\begin{align}
\iota_a(k,r)
&=
\sin(2kr)-2kr\cos(2kr),
\\
\iota_b(k,r)
&=
2kr\cos(2kr)
+
(2k^2r^2-1)\sin(2kr),
\\
\iota_c(k,r)
&=
-4kr\cos(2kr)
+
(2-2k^2r^2)\sin(2kr).
\end{align}
Their sine and cosine terms cancel pairwise, giving
\begin{equation}
\iota_b(k,r)+\iota_c(k,r)
=
\iota_a(k,r)
\end{equation}
identically. Equation~\eqref{eq:sumrule2} therefore holds for an
arbitrary isotropic monopole spectrum $f_0(k)$, rather than only for
a degenerate distribution.

Equations~\eqref{eq:sumrule1} and~\eqref{eq:sumrule2} reduce the
spin-dependent parity-conserving (SD-PC) kernels and several
combinations appearing below to a smaller set of independent
kernels. The remaining spin-dependent parity-violating (SD-PV)
structures are treated separately. These relations are algebraic and
do not depend on the particular background spectrum. To our
knowledge, they have not previously been used to analyze the bulk
Hartree limit of the background-induced neutrino interaction.

\subsection{Angular projection and the master Hartree formula}
\label{sec:master}

For a monochromatic directional background of number density $\Phi_0$,
energy $\Enu$, and direction $\khat$, the spin-independent (SI) potential
at $r\gg\Enu^{-1}$ is~\cite{Ghosh2023}
\begin{equation}
V^{\rm SI}(r,\theta)
=
-\frac{\GF^2\Phi_0\Enu}{\pi r}
g_V^1g_V^2\,F(\ell,\theta),
\label{eq:VSI}
\end{equation}
with
\begin{equation}
\begin{aligned}
F(\ell,\theta)
={}&
\cos^2\!\frac{\theta}{2}
\cos\!\left[(1-\cos\theta)\ell\right]
\\
&+
\sin^2\!\frac{\theta}{2}
\cos\!\left[(1+\cos\theta)\ell\right].
\end{aligned}
\label{eq:Fdirectional}
\end{equation}
Here $\ell\equiv\Enu r$, and $\theta$ is the angle between
$\mathbf r$ and $\khat$. At $\theta=0$ the phases vanish and
$V^{\rm SI}\propto1/r$. Away from this direction the potential
oscillates.

The directional structure simplifies considerably after the Hartree
orientation average.

\emph{Angular projection.}
We write the SI potential for an arbitrary ultrarelativistic occupation
\begin{equation}
f(\mathbf k)
=
f_\nu(\mathbf k)+f_{\bar\nu}(\mathbf k)
\end{equation}
as
\begin{equation}
V^{\rm SI}_f(\mathbf r)
=
\int\!\frac{d^3k}{(2\pi)^3}\,
f(\mathbf k)\,
\mathcal K_{\rm SI}
\!\left(r,k,\rhat\cdot\hat{\mathbf k}\right),
\label{eq:general_kernel}
\end{equation}
where the angular dependence of the scalar kernel enters only through
$\rhat\cdot\hat{\mathbf k}$. For fixed $\hat{\mathbf k}$, the integral
$\int d\Omega_{\rhat}\,\mathcal K_{\rm SI}$ cannot depend on the
direction $\hat{\mathbf k}$. By rotational invariance, it is therefore
the same angular integral obtained by fixing $\rhat$ and averaging over
$\hat{\mathbf k}$. Hence, identically in $r$,
\begin{equation}
\begin{aligned}
\frac{1}{4\pi}\int\!d\Omega_{\rhat}\,
V^{\rm SI}_f(r,\rhat)
&=
V^{\rm SI}_{f_0}(r),
\\
f_0(k)
&=
\frac{1}{4\pi}\int\!d\Omega_{\hat k}\,
f(k,\hat{\mathbf k}).
\end{aligned}
\label{eq:avgidentity}
\end{equation}
The orientation average therefore projects the background onto its
angular monopole. All higher angular multipoles vanish under this
projection.

Reference~\cite{Ghosh2023} numerically observed the corresponding
equivalence for a monochromatic directional beam.
Equation~\eqref{eq:avgidentity} extends this observation to an arbitrary
angular distribution and provides the starting point for the bulk
Hartree calculation. Once summed over pair orientations, a directional
beam is equivalent in the SI Hartree problem to an isotropic shell with
the same energy spectrum. The pair potential retains the beam direction,
whereas the homogeneous SI Hartree integral does not.

\emph{Master formula.}
For an isotropic background, Ref.~\cite{Ghosh2024} gives in the massless
limit
\begin{equation}
V^{\rm SI}_{\rm iso}(r)
=
-\frac{\GF^2g_V^1g_V^2}{4\pi^3r^5}
\mathcal J_a(r),
\label{eq:Viso}
\end{equation}
where
\begin{equation}
\mathcal J_a(r)
=
r\int_0^\infty\!dk\,f_0(k)
\left[
\sin(2kr)-2kr\cos(2kr)
\right].
\label{eq:Ja}
\end{equation}
This is the $\mathcal J$-factor formalism introduced in
Refs.~\cite{Horowitz1993,Ferrer1999}. The radial integral follows from
\begin{equation}
\frac{\sin(2kr)-2kr\cos(2kr)}{r^2}
=
-\frac{d}{dr}
\left[
\frac{\sin(2kr)}{r}
\right],
\label{eq:Ja_derivative}
\end{equation}
so that
\begin{equation}
\int_0^R\!dr\,
\frac{\sin(2kr)-2kr\cos(2kr)}{r^2}
=
2k-\frac{\sin(2kR)}{R}.
\label{eq:Ja_radial}
\end{equation}

To justify the interchange of the $r$ and $k$ integrations, we first
introduce a finite momentum cutoff $\Lambda$ as a mathematical
intermediate, perform the finite-$R$ radial integral exactly, and then
take $\Lambda\to\infty$. Since the form factor obtained below is
uniformly bounded and
$\int_0^\infty dk\,k f_0(k)<\infty$, dominated convergence applies.
No infrared regulator is required. Within the point-particle effective
kernel, the radial integral is also ultraviolet finite. The effective
description is not expected to remain valid down to $r=0$;
microscopic short-distance corrections are examined in
Sec.~\ref{sec:robustness}.

Writing
\begin{equation}
\mathcal H_{\rm SI}(R)
\equiv
\int_{|\mathbf r|<R}\!d^3r\,V^{\rm SI}(r),
\label{eq:Hlocaldef}
\end{equation}
the local homogeneous Hartree result is
\begin{equation}
\mathcal H_{\rm SI}(R)
=
-\frac{2\GF^2g_V^1g_V^2}{\pi^2}
\int_0^\infty\!dk\,k f_0(k)\,
\mathcal F(kR),
\label{eq:master}
\end{equation}
with
\begin{equation}
\mathcal F(x)
=
1-\frac{\sin(2x)}{2x},
\qquad
\mathcal F(x)\simeq
\begin{cases}
\dfrac{2}{3}x^2, & x\ll1,\\[5pt]
1+O(x^{-1}), & x\gg1.
\end{cases}
\label{eq:formfactor}
\end{equation}
Equation~\eqref{eq:master} is exact at finite $R$ for this local
homogeneous Hartree construction.

The form factor makes the short-wavelength condition
Eq.~\eqref{eq:validity} explicit mode by mode. For a spectral mode with
$kR\ll1$,
$\mathcal F(kR)\simeq\tfrac23(kR)^2$, so that its contribution to
$\mathcal H_{\rm SI}$ scales as $R^2$. For $kR\gg1$,
$\mathcal F(kR)\to1$, and the corresponding Hartree contribution
approaches an $R$-independent limit. For the monochromatic
free-streaming PNS benchmark, $\Enu R_\nu=1.8\times10^{18}$, and
therefore
\begin{equation}
\left|\mathcal F-1\right|
\le
\frac{1}{2\Enu R_\nu}
\simeq
3\times10^{-19}.
\label{eq:local_saturation}
\end{equation}

Taking $R\to\infty$ and assuming only that
\begin{equation}
M_1
\equiv
\int_0^\infty\!dk\,k f_0(k)
<
\infty,
\label{eq:M1}
\end{equation}
dominated convergence gives
\begin{equation}
\int d^3r\,V^{\rm SI}(r)
=
-\frac{2\GF^2g_V^1g_V^2}{\pi^2}\,M_1,
\label{eq:SIlimit}
\end{equation}
because $\mathcal F(kR)$ is uniformly bounded and
$k f_0(k)$ is integrable by assumption.

The construction above uses a local spherical Hartree cutoff. As a
finite-size check, consider instead a uniform sphere of radius $R$,
with constant neutron density and background spectrum. This geometry
can be treated exactly and allows us to verify that the saturation is
not an artifact of the local cutoff. For
\begin{equation}
\mathcal V_R=\frac{4\pi R^3}{3},
\end{equation}
the normalized overlap volume of two spheres whose centers are
separated by $r$ is
\begin{equation}
\begin{aligned}
C_R(r)
&\equiv
\frac{\mathcal V_{\rm ov}(r)}{\mathcal V_R}
\\
&=
1-\frac{3r}{4R}+\frac{r^3}{16R^3},
\qquad
0\le r\le2R.
\end{aligned}
\label{eq:sphere_window}
\end{equation}
Because $C_R(r)$ depends only on the separation magnitude, the angular
part of the finite-sphere convolution is exactly the projection in
Eq.~\eqref{eq:avgidentity}.

The interaction energy density for this uniform sphere is therefore
\begin{equation}
\mathcal E_{\rm sph}
\equiv
\frac{E_{\rm sph}}{\mathcal V_R}
=
\frac{n_n^2}{2}
\mathcal H_{\rm SI}^{\rm sph}(R),
\label{eq:sphere_energy}
\end{equation}
where
\begin{equation}
\mathcal H_{\rm SI}^{\rm sph}(R)
=
4\pi
\int_0^{2R}\!dr\,
r^2C_R(r)V_{\rm iso}^{\rm SI}(r).
\label{eq:sphere_hartree}
\end{equation}
Substituting Eqs.~\eqref{eq:Viso} and~\eqref{eq:Ja} gives
\begin{equation}
\begin{aligned}
\mathcal H_{\rm SI}^{\rm sph}(R)
={}&
-\frac{\GF^2g_V^1g_V^2}{\pi^2}
\int_0^\infty\!dk\,f_0(k)
\\
&\times
\int_0^{2R}\!dr\,
C_R(r)
\frac{\sin(2kr)-2kr\cos(2kr)}{r^2}.
\end{aligned}
\label{eq:sphere_double}
\end{equation}
The radial integral can be performed analytically, yielding
\begin{equation}
\begin{aligned}
\mathcal H_{\rm SI}^{\rm sph}(R)
={}&
-\frac{2\GF^2g_V^1g_V^2}{\pi^2}
\\
&\times
\int_0^\infty\!dk\,k f_0(k)\,
\mathcal F_{\rm sph}(kR),
\end{aligned}
\label{eq:sphere_master}
\end{equation}
where
\begin{equation}
\begin{aligned}
\mathcal F_{\rm sph}(x)
={}&
1-\frac{3\,\operatorname{Si}(4x)}{8x}
-\frac{3\cos(4x)}{32x^2}
\\
&+
\frac{3\sin(4x)}{128x^3},
\end{aligned}
\label{eq:sphere_formfactor}
\end{equation}
with
$\operatorname{Si}(z)\equiv
\int_0^z dt\,\sin t/t$.
The limiting forms are
\begin{equation}
\mathcal F_{\rm sph}(x)
=
\begin{cases}
\dfrac{8}{15}x^2+O(x^4), & x\ll1,\\[6pt]
1-\dfrac{3\pi}{16x}+O(x^{-3}), & x\gg1.
\end{cases}
\label{eq:sphere_limits}
\end{equation}

Thus the exact uniform-sphere geometry reproduces the expected
$R^2$ Hartree growth for long-wavelength modes, while in the
short-wavelength regime it approaches the same $R$-independent
Hartree limit as Eq.~\eqref{eq:SIlimit}. Finite-size geometry therefore
changes only boundary corrections and does not restore macroscopic
accumulation. For the free-streaming PNS benchmark, the leading
finite-sphere correction is
\begin{equation}
\frac{3\pi}{16\Enu R_\nu}
\simeq
3.3\times10^{-19}.
\label{eq:sphere_PNScorrection}
\end{equation}

At fixed density, the total interaction energy in the saturated regime
is still extensive,
$E_{\rm sph}\propto\mathcal V_R\propto R^3$.
What is absent is the additional $R^2$ growth of the Hartree
contribution per particle that would occur for an unsuppressed
long-range $1/r$ interaction.

At fixed monopole spectrum, the homogeneous SI Hartree integral is
independent of all higher angular multipoles of the background. Only
the monopole spectrum survives through the single moment $M_1$.
Directional beams, isotropic gases, partially anisotropic
semi-transparent distributions, and arbitrary mixtures of beam
directions are therefore all described by the same asymptotic result,
Eq.~\eqref{eq:SIlimit}.

Applying the same procedure to the spin-dependent kernels of
Ref.~\cite{Ghosh2024} also gives their radial moments. Writing the
radial integral in the dimensionless variable $y=2kr$, the
$\mathcal J_b$ kernel reduces to
\begin{equation}
\int_0^\infty\!dy\,
\left[
\frac{\cos y}{y}
-\frac{\sin y}{y^2}
+\frac{1}{2}\sin y
\right].
\label{eq:Jb_integral}
\end{equation}
The first two terms satisfy
\begin{equation}
\frac{\cos y}{y}
-\frac{\sin y}{y^2}
=
\frac{d}{dy}
\left(
\frac{\sin y}{y}
\right),
\label{eq:Jb_derivative}
\end{equation}
and therefore contribute
\begin{equation}
\left[
\frac{\sin y}{y}
\right]_0^\infty
=
-1.
\end{equation}
The remaining term gives $1/2$ in the Abel sense. The corresponding
moment is therefore $-M_1$. Together with the sum rule
Eq.~\eqref{eq:sumrule2}, this fixes
\begin{equation}
\int_0^\infty\!dr\,
\frac{\mathcal J_{a,b,c}(r)}{r^3}
=
\{2,\,-1,\,3\}\,M_1.
\label{eq:Jmoments}
\end{equation}
Unlike $\mathcal J_a$, the individual $\mathcal J_b$ and
$\mathcal J_c$ integrals are only conditionally convergent because
their integrands contain $\cos(2kr)/r$ tails. They are therefore
understood in the Abel sense, as the $\eta\to0^+$ limit obtained after
introducing an $e^{-\eta r}$ regulator. The SI result
Eq.~\eqref{eq:SIlimit} requires no such prescription.

\subsection{The two proto-neutron-star regimes}

The free-streaming and trapped-core limits follow from different choices
of the monopole spectrum $f_0(k)$.

\emph{Free-streaming region:}
For a monochromatic directional background we write
\begin{equation}
f_{\rm dir}(\mathbf k)
=
(2\pi)^3\Phi_0\,
\delta^3(\mathbf k-\mathbf k_0),
\qquad
|\mathbf k_0|=\Enu ,
\end{equation}
whose monopole is
\begin{equation}
f_0(k)
=
\frac{2\pi^2\Phi_0}{k^2}\,
\delta(k-\Enu).
\end{equation}
It is normalized according to
\begin{equation}
\int\!\frac{d^3k}{(2\pi)^3}\,f_0(k)
=
\Phi_0 .
\end{equation}
The corresponding first moment is
\begin{equation}
M_1
=
\frac{2\pi^2\Phi_0}{\Enu},
\end{equation}
and Eq.~\eqref{eq:SIlimit} gives
\begin{equation}
\int\! d^3r\,V^{\rm SI}
=
-\frac{4\,g_V^1g_V^2\,\GF^2\,\Phi_0}{\Enu}
\label{eq:beam}
\end{equation}
in the $R\to\infty$ saturated homogeneous Hartree limit.

A Coulomb ansatz $V=\alpha_{\rm eff}/r$ would instead give
$2\pi\alpha_{\rm eff}R^2$. Matching $\alpha_{\rm eff}$ to the forward
$1/r$ coefficient of Eq.~\eqref{eq:VSI}, the saturated Hartree result
is smaller by
\begin{equation}
\frac{2}{(\Enu R)^2}
\simeq
6\times10^{-37}
\end{equation}
at $R=R_\nu$. The reduction is therefore not a screening effect,
because no exponential cutoff has been introduced.

\emph{Degenerate core:}
For a zero-temperature degenerate neutrino gas,
\begin{equation}
f_0(k)=\Theta(\munu-k),
\end{equation}
so that
\begin{equation}
M_1=\frac{\munu^2}{2}.
\end{equation}
Equation~\eqref{eq:SIlimit} then gives
\begin{equation}
\mathcal E_{2\nu}
=
\frac12\nn^2\!\int\! d^3r\,V^{\rm SI}
=
-\frac{\GF^2 g_V^1g_V^2}{2\pi^2}\,
\nn^2\munu^2 .
\label{eq:core}
\end{equation}
Comparing with the free-gas neutron degeneracy pressure
\begin{equation}
P_{\rm deg}
=
\frac{k_F^5}{15\pi^2m_n},
\qquad
k_F=(3\pi^2 n_n)^{1/3},
\end{equation}
one obtains
\begin{equation}
\frac{\mathcal E_{2\nu}}{P_{\rm deg}}
=
-\frac{5}{6\pi^4}\,
\GF^2 g_V^1g_V^2\,
m_n k_F \munu^2 .
\label{eq:coreratio}
\end{equation}

\emph{Thermal core:}
For an ultrarelativistic equilibrium neutrino gas, with
$f_0=f_\nu+f_{\bar\nu}$ and
$\mu_{\bar\nu}=-\mu_\nu$, one finds
\begin{equation}
M_1
=
\frac{\munu^2}{2}
+
\frac{\pi^2T^2}{6}.
\end{equation}
Equation~\eqref{eq:core} is therefore rescaled by
\begin{equation}
1+\frac{\pi^2T^2}{3\munu^2}.
\end{equation}
This gives a $0.8\%$ correction at $T=10\,\mathrm{MeV}$ and
$7.4\%$ at $T=30\,\mathrm{MeV}$. The degenerate approximation is
therefore accurate for the benchmarks used here, while the
non-accumulation argument itself does not rely on degeneracy. 

No separate angular treatment is required for the SI contribution in
the semi-transparent region. For arbitrary local angular anisotropy,
the same master formula applies, and the magnitude is fixed solely by
the local monopole moment $M_1$. Anisotropy therefore cannot restore SI
accumulation.

\emph{Nonuniformity of the directional asymptotic form:}
Equation~\eqref{eq:VSI} must not be integrated directly over
$d\Omega_{\rhat}$. It is derived at fixed $\theta$ for $\ell\gg1$,
and this limit is not uniform in angle. The non-oscillating part of
$F$ arises from the forward endpoint, within a cone
$\theta\lesssim\ell^{-1/2}$ that shrinks as $\ell$ increases. The
relevant local expansion parameter there is therefore
$\theta^2\ell$, rather than $\ell$ itself.

A term-by-term angular integration of the large-$\ell$ directional
form formally yields
\begin{equation}
\frac{2\pi\left(1-\cos2\ell\right)}{\ell^2}.
\end{equation}
Because this quantity is nonnegative, its subsequent radial integration
would produce a spurious logarithmic growth. This is an artifact of
interchanging the angular integral with a nonuniform large-$\ell$
expansion. Equation~\eqref{eq:avgidentity} avoids this problem by
performing the angular projection before invoking the directional
asymptotic form.

Accordingly, the directional $1/r$ enhancement should not be
interpreted as an isotropic interaction range. The $1/r$ behavior
persists only while
\begin{equation}
(1-\cos\theta)\Enu r\lesssim1.
\end{equation}
For small angles,
\begin{equation}
(1-\cos\theta)\Enu r
\simeq
\frac{\theta^2\Enu r}{2},
\end{equation}
so the condition defines a coherence cone
\begin{equation}
\theta_{\rm coh}(r)
=
\sqrt{\frac{2}{\Enu r}}.
\label{eq:cone}
\end{equation}
This cone subtends a fraction
$\sim\theta_{\rm coh}^2/4$ of all orientations, corresponding to
$8\times10^{-9}$ at $r=1\,\mu\mathrm{m}$ and
$8\times10^{-13}$ at $r=1\,\mathrm{cm}$ for
$\Enu=12\,\mathrm{MeV}$.

Outside the cone the oscillation wavelength is
\begin{equation}
\lambda_{\rm osc}
=
\frac{2\pi}
{(1-\cos\theta)\Enu},
\end{equation}
which gives $0.7\,\mathrm{nm}$ at $\theta=1^\circ$ and
$0.10\,\mathrm{pm}$ at $\theta=90^\circ$. This narrow angular support
is a property of the pairwise potential. Equation~\eqref{eq:avgidentity}
shows that it does not survive the SI Hartree angular average.

\subsection{Spin-dependent and parity-violating sectors}
\label{sec:spin}

The full SM neutrino force contains, in addition to the SI part, a
spin-dependent parity-conserving (SD-PC) and a spin-dependent
parity-violating (SD-PV) part~\cite{Ghosh2024}. In a background, the
SD-PV terms are not subject to the same velocity suppression as in
vacuum, since the background breaks Lorentz invariance. One might
therefore expect them to evade the cancellations found in the SI
sector. They do not for the PNS configurations considered here. In
what follows, $\bsig_{1,2}$ denote the Pauli spin operators of the two
external fermions, andand $\hat{\mathbf k}_0$ denotes the direction of the background-neutrino beam.

\paragraph{SD-PC.}
The projection identity, Eq.~\eqref{eq:avgidentity}, does not extend
directly to the SD-PC sector. With spins present one can construct
structures such as
$(\bsig_1\!\cdot\!\khat)(\bsig_2\!\cdot\!\khat)$, so $\khat$
survives the $\rhat$ average as a preferred direction and anisotropic
multipoles of the background can propagate through the spin tensors.
The directional SD-PC contribution must therefore be evaluated
separately.

Reference~\cite{Ghosh2024} writes the directional SD-PC potential as
\begin{equation}
\begin{aligned}
V^{\rm SD\text{-}PC}_{\rm bkg}
=
-\frac{\GF^2\Phi_0\Enu}{2\pi^3}
\Big\{
&g_A^1g_A^2
\Big[
(\bsig_1\!\cdot\!\bsig_2)\mathcal I_a
\\
&\qquad+
(\bsig_1\!\cdot\!\nabla)
(\bsig_2\!\cdot\!\nabla)\mathcal I_b
\Big]
\\
&-2g_V^2g_A^1
\Big[
(\bsig_1\!\cdot\!\khat)\mathcal I^{\rm PV}
\\
&\qquad+
(\bsig_1\!\cdot\!\nabla)\mathcal I_c
\Big]
-(1\leftrightarrow2)
\Big\}.
\end{aligned}
\label{eq:sdpcdir}
\end{equation}
Following the notation of Ref.~\cite{Ghosh2024}, the
$1\leftrightarrow2$ operation applies to the preceding mixed
$g_Vg_A$ contribution; the $g_A^1g_A^2$ bracket is already symmetric.

First, the sum rule Eq.~\eqref{eq:sumrule1} reduces $\mathcal I_a$ to
$\mathcal I^{\rm SI}$ and $\mathcal I^{\rm PV}$. The momentum-space
route gives both angular averages in closed form:
\begin{equation}
\int\!d\Omega_{\rhat}\,
\mathcal I^{\rm SI}
=
\frac{2\pi^3}{r}
\frac{\sin(2\ell)-2\ell\cos(2\ell)}{\ell^3},
\label{eq:ISIangular}
\end{equation}
and
\begin{equation}
\int\!d\Omega_{\rhat}\,
\mathcal I^{\rm PV}
=
\frac{4\pi^3}{r}
\frac{\sin(2\ell)}{\ell}.
\label{eq:IPVangular}
\end{equation}
Both are purely oscillatory with no constant part, so
\begin{equation}
\begin{aligned}
\int d\Omega_{\rhat}\,\mathcal I_a
={}&
2\int d\Omega_{\rhat}\,\mathcal I^{\rm PV}
\\
&-
\int d\Omega_{\rhat}\,\mathcal I^{\rm SI}
\end{aligned}
\end{equation}
is manifestly oscillatory, and the corresponding volume-weighted
Hartree radial integrals are bounded.

With
\begin{equation}
\mathcal I_b
=
\int d^3q\,
e^{i\mathbf q\cdot\mathbf r}
\frac{1+\xi^2}
{\rho^2(\rho^2-4\Enu^2\xi^2)}
\end{equation}
from Ref.~\cite{Ghosh2024}, each gradient brings down
$i\mathbf q$, so
\begin{equation}
\begin{aligned}
&(\bsig_1\!\cdot\!\nabla)
(\bsig_2\!\cdot\!\nabla)\mathcal I_b
\\
&\qquad\longleftrightarrow
-(\bsig_1\!\cdot\!\hat{\mathbf q})
(\bsig_2\!\cdot\!\hat{\mathbf q})
\frac{1+\xi^2}
{\rho^2-4\Enu^2\xi^2}.
\end{aligned}
\label{eq:Ibmomentum}
\end{equation}
The tensor term is simplest in momentum space. Each derivative supplies
a factor $i\mathbf q$, so the resulting factor $\rho^2$ cancels the
explicit $1/\rho^2$ in $\mathcal I_b$. The tensor term therefore has
the same propagator denominator as the SI term and receives no
additional long-wavelength suppression.

Its bulk sum is nonetheless bounded. Inserting the spherical Hartree
window
\begin{equation}
W_R(\rho)
=
4\pi
\frac{\sin(\rho R)-\rho R\cos(\rho R)}{\rho^3}
\end{equation}
and using axial symmetry about $\khat$, the tensor contribution
decomposes as
\begin{equation}
T_{ij}
=
T_\perp\Delta_{ij}
+
T_\parallel\hat k_{0i}\hat k_{0j},
\qquad
\Delta_{ij}
=
\delta_{ij}-\hat k_{0i}\hat k_{0j}.
\end{equation}
Each coefficient is proportional, up to an $R$-independent factor, to
\begin{equation}
\mathcal T_A(L)
=
\int_0^\infty\!da\,
\frac{\sin(2La)-2La\cos(2La)}{a}\,
A(a),
\qquad
L=\Enu R,
\label{eq:Tij}
\end{equation}
with $A(a)=\alpha(a)$ for $T_\perp$ and
$A(a)=\beta(a)$ for $T_\parallel$.

Writing
\begin{equation}
a=\frac{\rho}{2\Enu},
\qquad
\Lambda_0(a)
=
\frac{1}{a}
\ln\left|
\frac{1+a}{1-a}
\right|,
\end{equation}
the principal-value angular functions are
\begin{equation}
\begin{aligned}
\alpha(a)
&=
\frac{2}{3}
+2a^2
+(1-a^4)\Lambda_0(a),
\\
\beta(a)
&=
-\frac{8}{3}
-2a^2
+a^2(a^2+1)\Lambda_0(a).
\end{aligned}
\label{eq:alphabeta}
\end{equation}
At the origin,
$A(a)-A(0)=O(a^2)$, with
$\alpha(0)=8/3$ and $\beta(0)=-8/3$; both functions fall as
$O(a^{-2})$ for $a\to\infty$. Their only interior nonanalyticity is
at $a=1$, where $\alpha(a)$ remains finite with an
$(a-1)\ln|a-1|$ cusp, while
\begin{equation}
\beta(a)
=
-2\ln|a-1|+O(1)
\end{equation}
has an integrable logarithmic singularity.

Write
\begin{equation}
\mathcal T_A(L)
=
S_A(L)-2L\,C_A(L),
\end{equation}
where
\begin{equation}
\begin{aligned}
S_A(L)
&=
\int_0^\infty\!da\,
\frac{A(a)}{a}\sin(2La),
\\
C_A(L)
&=
\int_0^\infty\!da\,
A(a)\cos(2La).
\end{aligned}
\end{equation}
For the sine transform, the origin value must be subtracted with an
integrable profile. Writing
\begin{equation}
\frac{A(a)}{a}
=
A(0)\frac{e^{-a}}{a}
+
\frac{A(a)-A(0)e^{-a}}{a},
\end{equation}
the second term is $O(1)$ at the origin, $O(a^{-3})$ at infinity,
and integrable at $a=1$, while
\begin{equation}
\int_0^\infty\!da\,
e^{-a}\frac{\sin(2La)}{a}
=
\arctan(2L).
\end{equation}
The Riemann--Lebesgue lemma then gives
\begin{equation}
S_A(L)
=
A(0)\arctan(2L)+o(1)
\longrightarrow
\frac{\pi}{2}A(0).
\end{equation}

For the cosine transform, integration by parts eliminates both boundary
terms and gives
\begin{equation}
2L\,C_A(L)
=
-\int_0^\infty\!da\,
A'(a)\sin(2La).
\end{equation}
Since $\alpha'\in L^1(0,\infty)$, the Riemann--Lebesgue lemma yields
$2L\,C_\alpha(L)=o(1)$ and therefore
\begin{equation}
\mathcal T_\alpha(L)
\longrightarrow
\frac{4\pi}{3}.
\end{equation}

For $\beta(a)$, the logarithm implies locally
$\beta'(a)\sim-2/(a-1)$. The subtraction must be localized, because
$\beta'(a)=O(a^{-3})$ at infinity whereas $(a-1)^{-1}$ is not
integrable there. Introducing a smooth
$\zeta\in C_c^\infty(0,\infty)$ equal to unity near $a=1$,
\begin{equation}
\beta'(a)
=
-2\zeta(a)\,
\operatorname{PV}\frac{1}{a-1}
+b(a),
\qquad
b\in L^1(0,\infty),
\end{equation}
and
\begin{equation}
\operatorname{PV}
\int_0^\infty\!da\,
\zeta(a)
\frac{\sin(2La)}{a-1}
=
\pi\cos(2L)+o(1).
\end{equation}
Hence
\begin{equation}
\mathcal T_\beta(L)
=
-\frac{4\pi}{3}
-2\pi\cos(2L)
+o(1),
\label{eq:Tbeta}
\end{equation}
and both tensor projections are $O(1)$ as $L\to\infty$. The
transverse component approaches a constant, while the longitudinal
one remains bounded and oscillatory. The directional tensor
contribution therefore does not accumulate.

The mixed $g_Vg_A$ contribution obeys an exact angular identity. With
\begin{equation}
\mathcal I^{\rm PV}
=
\int d^3q\,
\frac{e^{i\mathbf q\cdot\mathbf r}}
{\rho^2-4\Enu^2\xi^2}
\end{equation}
and
\begin{equation}
\mathcal I_c
=
i\int d^3q\,
e^{i\mathbf q\cdot\mathbf r}
\frac{\xi}
{\rho(\rho^2-4\Enu^2\xi^2)},
\end{equation}
one has
\begin{equation}
\nabla\mathcal I_c
=
-\int d^3q\,
e^{i\mathbf q\cdot\mathbf r}
\frac{\hat{\mathbf q}\,\xi}
{\rho^2-4\Enu^2\xi^2}.
\end{equation}
The bracket therefore has numerator
$(\bsig\!\cdot\!\khat)
-(\bsig\!\cdot\!\hat{\mathbf q})\xi$.
Since
\begin{equation}
\int d\Omega_{\rhat}\,
e^{i\mathbf q\cdot\mathbf r}
=
4\pi j_0(\rho r),
\end{equation}
where $j_0$ is the spherical Bessel function of order zero, the
surviving angular average satisfies
\begin{equation}
\int d\Omega_{\hat{\mathbf q}}\,
\hat{\mathbf q}\,\xi\,h(\xi)
=
\khat
\int d\Omega_{\hat{\mathbf q}}\,
\xi^2h(\xi)
\end{equation}
for any function $h$. The numerator left over is $1-\xi^2$, precisely
the SI numerator. Hence
\begin{equation}
\begin{aligned}
&\int\!d\Omega_{\rhat}
\left[
(\bsig\!\cdot\!\khat)\mathcal I^{\rm PV}
+
(\bsig\!\cdot\!\nabla)\mathcal I_c
\right]
\\
&\qquad=
(\bsig\!\cdot\!\khat)
\int\!d\Omega_{\rhat}\,
\mathcal I^{\rm SI}
\end{aligned}
\label{eq:VAidentity}
\end{equation}
exactly, with no large-$\ell$ approximation. The mixed contribution
therefore inherits the bounded, saturating behavior of the SI angular
kernel.

All three directional SD-PC structures are therefore bounded in the
homogeneous Hartree configuration considered here, although their spin
weighting differs. We denote the spin-polarization fraction by $P$,
with
\begin{equation}
\langle\bsig\rangle
=
P\,\hat{\mathbf n},
\end{equation}
where $\hat{\mathbf n}$ is the polarization axis. Using the nuclear
magneton $\mu_N$ as the characteristic magnetic-moment scale gives
$P\sim\mu_NB/T\sim3\times10^{-4}$ for
$B=10^{15}\,\mathrm{G}$ and $T=10\,\mathrm{MeV}$.
For a neutron pair, the $g_A^1g_A^2$ structures are therefore
quadratic in the polarization and carry a suppression of order
$P^2\sim10^{-7}$ relative to the corresponding unpolarized spin
scale. The mixed $g_Vg_A$ structure is linear in $P$ and is
parametrically larger, but remains negligible relative to an SI
contribution that is already suppressed at the $10^{-15}$ level.

In an isotropic background, the SD-PC contribution vanishes in the
infinite homogeneous Hartree limit, while the finite-$R$ remainder is
a non-accumulating boundary term. The potential contains the two
structures
$(\bsig_1\!\cdot\!\bsig_2)\mathcal J_b$ and
$(\bsig_1\!\cdot\!\rhat)(\bsig_2\!\cdot\!\rhat)\mathcal J_c$.
Using
\begin{equation}
\int d\Omega_{\rhat}\,
\hat r_i\hat r_j
=
\frac{4\pi}{3}\delta_{ij},
\end{equation}
the second reduces to
$\frac{4\pi}{3}
(\bsig_1\!\cdot\!\bsig_2)\mathcal J_c$, so the angular integral
leaves
$\mathcal J_b+\tfrac13\mathcal J_c$.
The radial moments Eq.~\eqref{eq:Jmoments} then give
\begin{equation}
\int_0^\infty\!dr\,
\frac{\mathcal J_b+\tfrac13\mathcal J_c}{r^3}
=
-M_1+\frac13(3M_1)
=
0.
\label{eq:sdpczero}
\end{equation}
The cancellation is exact for any isotropic ultrarelativistic
distribution with finite $M_1$ in this infinite-volume
Abel-regularized sense.

For a finite homogeneous region the cancellation is approached through
a non-accumulating boundary term rather than being identically zero.
For the filled Fermi sphere, writing
\begin{equation}
X=\munu R,
\end{equation}
direct integration gives
\begin{equation}
\begin{aligned}
\int_0^R\!dr\,
\frac{\mathcal J_b+\tfrac13\mathcal J_c}{r^3}
=
\frac{\munu^2}{3X^2}
\Big[
1-\cos(2X)-X\sin(2X)
\Big].
\end{aligned}
\label{eq:sdpcfinite}
\end{equation}
The finite-$R$ remainder is therefore
$O[(\munu R)^{-1}]$ relative to the SI Hartree term and vanishes in
the $R\to\infty$ limit of Eq.~\eqref{eq:sdpczero}. At the PNS
benchmark, $\munu R\simeq3.0\times10^{19}$, the finite-size correction
at the level of the kernel is of order $10^{-20}$ before polarization
factors.

Like the individual moments, the infinite-volume surviving combination
is only conditionally convergent. For the filled Fermi sphere,
\begin{equation}
\frac{\mathcal J_b+\tfrac13\mathcal J_c}{r^3}
=
-\frac23\munu^2
\frac{\cos(2\munu r)}{r}
+
O(r^{-2})
\end{equation}
at large $r$, so Eq.~\eqref{eq:sdpczero} is understood in the same Abel
sense as the $\mathcal J_{b,c}$ moments of Eq.~\eqref{eq:Jmoments}.
Consequently, at the uncorrelated Hartree level the core receives no
SD-PC contribution in the infinite homogeneous limit, while its
finite-size remainder is suppressed by $O[(\munu R)^{-1}]$ and is
completely negligible at PNS scales.
Section~\ref{sec:robustness} examines how short-range correlations
modify this cancellation.

\paragraph{SD-PV.}
The leading SD-PV potential contains two spin structures
~\cite{Ghosh2024}:
\begin{equation}
\begin{aligned}
\Big[
&g_A^1g_A^2
(\bsig_1\times\bsig_2)
\\
&+
\left(
g_V^1g_A^2\bsig_2
-
g_V^2g_A^1\bsig_1
\right)\times\khat
\Big]\cdot\rhat .
\end{aligned}
\end{equation}
Both give zero Hartree contribution for the collinearly polarized
directional mean-field state considered here. The first term is
proportional to the cross product of the two spin expectation values.
In the factorized mean-field state,
\begin{equation}
\begin{aligned}
\langle\bsig_1\times\bsig_2\rangle
&=
\langle\bsig_1\rangle
\times
\langle\bsig_2\rangle
\\
&=
P^2
(\hat{\mathbf n}\times\hat{\mathbf n})
=
0.
\end{aligned}
\end{equation}
The operator itself does not vanish; rather, its ensemble expectation
value vanishes in the collinearly polarized factorized mean-field
state.

The second structure vanishes at $\theta=0$ by the triple product,
since there $\rhat\parallel\khat$ and
$(\bsig\times\khat)\cdot\khat=0$. More generally, its Hartree integral
vanishes exactly in the directional configuration considered here.
The scalar kernel depends on $\rhat$ only through
$\rhat\cdot\khat$, so azimuthal symmetry implies
\begin{equation}
\int d\Omega_{\rhat}\,
\rhat\,G(\rhat\cdot\khat)
\parallel
\khat,
\end{equation}
whereas
$(\hat{\mathbf n}\times\khat)\perp\khat$.

For an isotropic background, the leading SD-PV Hartree contribution
vanishes by angular symmetry without any assumption about the
polarization. Reference~\cite{Ghosh2024} gives the leading isotropic
SD-PV potential as
\begin{equation}
V^{\rm SD-PV}_{\rm bkg,0}
\propto
\rhat\cdot(\bsig_1\times\bsig_2)\,
\mathcal J_d(r),
\label{eq:sdpviso}
\end{equation}
and since
$\int d\Omega_{\rhat}\,\rhat=0$, its Hartree contribution vanishes
identically.

Independently of these cancellations, the coherent SD-PV contribution
is further suppressed relative to the forward SI $1/r$ term by
$O[(\Enu r)^{-1}]$; none of the conclusions above relies on this
additional suppression.

\paragraph{Velocity-suppressed SD-PV.}
The subleading parity-violating terms~\cite{Ghosh2024} involve the
reduced velocities
\begin{equation}
\widetilde{\mathbf v}_{1,2}
=
\mathbf v_{1,2}
\pm
\frac{i\,\bsig_{1,2}\times\mathbf q}{2m_{1,2}},
\label{eq:reducedv}
\end{equation}
where $\mathbf v_{1,2}$ and $m_{1,2}$ are the ordinary velocities and
masses of the two external fermions. Both pieces must therefore be
addressed. The ordinary velocity contribution cannot be neglected
solely on the basis of $v\ll1$, since in the core
$k_F/m_n\simeq0.44$, corresponding to
\begin{equation}
\frac{v_F}{c}
=
\frac{k_F}{\sqrt{k_F^2+m_n^2}}
\simeq
0.40,
\end{equation}
while thermal velocities in the free-streaming region are
$v/c\simeq0.13$--$0.18$.

The ordinary-velocity pieces are linear in $\mathbf v_i$. In the local
rest frame of an isotropic equilibrium medium,
$\langle\mathbf v\rangle=0$, so these terms vanish upon averaging,
independent of the magnitude of $v_F$.

The $\mathbf q/m$ terms reduce either to identically vanishing
antisymmetric contractions or to boundary terms. After
$\mathbf q\to-i\nabla$, they generate two kinds of contribution.
Terms in which the resulting gradient is contracted with a second
gradient, of the form
$(\bsig\times\nabla)\cdot\nabla$, vanish identically by antisymmetry:
\begin{equation}
\epsilon_{ijk}\partial_j\partial_k=0,
\end{equation}
where $\epsilon_{ijk}$ is the Levi-Civita symbol. Terms linear in a
single gradient reduce in the Hartree integral to a surface term,
\begin{equation}
\int_V d^3r\,\partial_i\Psi
=
\oint_{\partial V}dS_i\,\Psi.
\end{equation}
In the isotropic case this vanishes by angular symmetry. In the
directional case the surface vector is parallel to $\khat$; terms of
the form $(\bsig\times\nabla)\cdot\khat$ vanish because
$\epsilon_{ijk}\hat k_{0j}\hat k_{0k}=0$, while structures reducing to
$(\bsig_1\times\bsig_2)\cdot\khat$ vanish in the collinearly polarized
mean-field state considered above. These derivative pieces therefore
generate no accumulating Hartree contribution.

A nonzero ordinary-velocity contribution requires a bulk flow
$\mathbf u$, as supplied by convection or rotation. Denoting by
$\mathbf u_{1,2}$ the local bulk velocities associated with the two
external fermions, terms multiplying $\mathcal I_b$ and
$\mathcal I_c$ retain explicit spatial derivatives, for example
\begin{equation}
\begin{aligned}
&(\bsig_1\!\cdot\!\nabla)
(\mathbf u_2\!\cdot\!\nabla)\mathcal I_b,
\\
&(\mathbf u_1\!\cdot\!\nabla)\mathcal I_c.
\end{aligned}
\end{equation}
Their Hartree integrals reduce to boundary structures such as
\begin{equation}
\int_V d^3r\,
\partial_i\partial_j\mathcal I_b
=
\oint_{\partial V}
dS_i\,\partial_j\mathcal I_b,
\end{equation}
and
\begin{equation}
\int_V d^3r\,
\partial_i\mathcal I_c
=
\oint_{\partial V}
dS_i\,\mathcal I_c.
\end{equation}

As in the SD-PC case, the explicit powers of $\mathbf q$ brought down
by the gradients cancel corresponding inverse powers carried by the
kernels, so no additional gradient-scale suppression is assumed.
Instead, the $\mathcal I_b$ tensor structure is bounded by
Eq.~\eqref{eq:Tij}, and the $\mathcal I_c$ gradient combines with
$\mathcal I^{\rm PV}$ through the exact identity
Eq.~\eqref{eq:VAidentity}. The remaining bulk-flow terms built on
$\mathcal I^{\rm PV}$ have purely oscillatory angular averages, while
those built on $\mathcal I^{\rm SI}$ and $\mathcal I_a$ are bounded
for the same reason and carry additional polarization and flow
factors. For representative values
$P\sim3\times10^{-4}$ and $u/c\sim0.03$,
\begin{equation}
P\frac{u}{c}
\sim
10^{-5}.
\end{equation}
Thus the spin-dependent parity-violating sectors generate no
accumulating Hartree contribution under the PNS conditions considered.

\section{PNS Numerical Benchmarks}
\label{sec:numbers}

For the free-streaming region we adopt
\[
R_\nu=30\,\mathrm{km},
\qquad
L_\nu=3\times10^{52}\,\mathrm{erg\,s^{-1}},
\qquad
\langle E_\nu\rangle=12\,\mathrm{MeV},
\]
where $L_\nu$ is summed over neutrino species, together with a mass
density
$\rho_m=10^{11}\,\mathrm{g\,cm^{-3}}$ at the neutrinosphere.
The corresponding free-streaming neutrino number density is
\begin{equation}
n_\nu
=
\frac{L_\nu}
{4\pi R_\nu^2\langle E_\nu\rangle c}
=
4.6\times10^{32}\,\mathrm{cm^{-3}},
\end{equation}
while the baryon-density normalization gives
\begin{equation}
n_b
\simeq
\frac{\rho_m}{m_n}
=
6.0\times10^{34}\,\mathrm{cm^{-3}}.
\end{equation}
For the neutron-only benchmark we set $n_n=n_b$. If
$Y_e\equiv n_e/n_b=0.3$, the neutron density is instead
$n_n\simeq(1-Y_e)n_b\simeq0.7n_b$, which only reduces the
neutron-pair contribution and does not affect the conclusions.

For the monochromatic free-streaming benchmark used below, we identify
\begin{equation}
\Phi_0=n_\nu,
\qquad
E_\nu=\langle E_\nu\rangle=12\,\mathrm{MeV}.
\end{equation}

For the trapped core we take
\begin{equation}
n_n=3\times10^{38}\,\mathrm{cm^{-3}},
\qquad
\mu_\nu=200\,\mathrm{MeV}.
\end{equation}
Gravitational comparisons use
$M=1.4M_\odot$ and $R=R_\nu$. For reference, we define
\begin{equation}
|V_{\rm grav}(r)|
=
\frac{Gm_n^2}{r},
\qquad
|U_{\rm grav}|
=
\frac{GMm_n}{R_\nu}.
\end{equation}

For the force-density comparison we use the characteristic local scale
height $H=R_\nu/2$ and define
\begin{equation}
\mathfrak f_{2\nu}
=
n_n\frac{|U_{2\nu}|}{H},
\qquad
\mathfrak f_{\rm grav}
=
\rho_m\frac{GM}{R_\nu^2}.
\end{equation}
Here $\mathfrak f$ denotes force per unit volume, distinguishing these
quantities from the phase-space distributions $f(\mathbf k)$ used
above.

For the interaction-pressure comparison we define the \emph{partial}
interaction pressure associated with neutron compression at fixed
neutrino-background parameters:
\begin{equation}
P_{2\nu}^{(n)}
=
n_n
\left(
\frac{\partial\mathcal E_{2\nu}}
{\partial n_n}
\right)_{\rm bkg}
-
\mathcal E_{2\nu}.
\end{equation}
The superscript $(n)$ indicates that the derivative is taken with
respect to the neutron density while the background parameters are
held fixed. Since
$\mathcal E_{2\nu}\propto n_n^2$ at fixed background,
\begin{equation}
P_{2\nu}^{(n)}
=
\mathcal E_{2\nu}.
\label{eq:partialpressure}
\end{equation}
This is the appropriate partial Hartree interaction pressure for the
benchmark comparison. It is not a complete multicomponent PNS
equation of state, in which the neutrino background and composition
would co-evolve under compression.

For the free-streaming benchmark we compare this interaction pressure
with the ultrarelativistic degenerate-electron pressure at $Y_e=0.3$,
\begin{equation}
P_{e,\rm deg}
=
1.2\times10^{29}\,
\mathrm{dyn\,cm^{-2}}.
\end{equation}

\begin{table}[h!]
\caption{Neutrino-force benchmarks in a proto-neutron star. The large
forward pair enhancement does not persist in the corresponding bulk
Hartree quantities.}
\label{tab:results}
\begin{ruledtabular}
\begin{tabular}{lc}
Quantity & Value \\
\hline
\multicolumn{2}{l}{\emph{Free-streaming region (directional)}}\\
$|V^{\rm SI}(\theta=0)|/|V_{\rm grav}|$ (pair)
& $7.8\times10^{16}$ \\
one-body Hartree potential $U_{2\nu}$
& $-1.8\times10^{-14}\,\mathrm{eV}$ \\
$|U_{2\nu}|/|U_{\rm grav}|$
& $2.9\times10^{-22}$ \\
$|P_{2\nu}^{(n)}|/P_{e,\rm deg}$
& $7.4\times10^{-21}$ \\
$\mathfrak f_{2\nu}/\mathfrak f_{\rm grav}$
& $5.5\times10^{-22}$ \\
\hline
\multicolumn{2}{l}{\emph{Degenerate core (isotropic)}}\\
$\mathcal E_{2\nu}$
& $-7.6\times10^{19}\,\mathrm{erg\,cm^{-3}}$ \\
$|\mathcal E_{2\nu}|/P_{\rm deg}$
& $4.5\times10^{-15}$ \\
inverse neutrino momentum scale $1/\mu_\nu$
& $0.99\,\mathrm{fm}$ \\
\end{tabular}
\end{ruledtabular}
\end{table}

Table~\ref{tab:results} summarizes the benchmark results. The forward
pair potential exceeds the neutron--neutron gravitational potential by
a factor $7.8\times10^{16}$, whereas the corresponding one-body
Hartree potential is only $2.9\times10^{-22}$ of the stellar
gravitational potential per neutron. The ratio of these two
dimensionless comparisons is approximately $4\times10^{-39}$. This
large difference illustrates that the forward pair enhancement is not
representative of the orientation-summed bulk Hartree interaction.

These parameter choices are representative benchmarks for the
deleptonization epoch rather than outputs of a self-consistent
time-dependent PNS simulation. Such modeling would be required for a
precision prediction, but is not needed to establish the hierarchy
relevant here, which spans many orders of magnitude.

The semi-transparent region near the neutrinosphere requires no
separate angular treatment in the SI sector. At the local homogeneous
Hartree level, Eq.~\eqref{eq:SIlimit} fixes the bulk contribution
solely through the monopole moment $M_1$.

Equation~\eqref{eq:core} depends only on $n_n$ and $\mu_\nu$ within
the uncorrelated neutron Hartree benchmark and contains no explicit
EOS model beyond these inputs. At fixed $n_n$, a larger physical
pressure only decreases the fractional importance of the interaction.
The benchmark is neutron-only; a self-consistent multicomponent
treatment including electrons, protons, flavor-dependent neutrino
populations, composition constraints, and correlations would be
required for precision EOS calculations. Order-unity modifications
from composition or correlations cannot compensate the benchmark
suppression
$|\mathcal E_{2\nu}|/P_{\rm deg}=4.5\times10^{-15}$.

For the background-neutrino momenta relevant here,
$k\gg m_\nu$, so neutrino-mass corrections and Dirac--Majorana
differences are parametrically negligible for the background-induced
terms considered.

Finally, additional matter insertions on the hard
background-neutrino modes are perturbative. A useful expansion
parameter is
\begin{equation}
\epsilon_{\rm mat}
\equiv
\frac{\GF n_n}{k_{\rm bkg}},
\end{equation}
where $k_{\rm bkg}=E_\nu$ in the free-streaming benchmark and
$k_{\rm bkg}\sim\mu_\nu$ in the degenerate core. Numerically,
\begin{equation}
\epsilon_{\rm mat}
\simeq
\begin{cases}
4.5\times10^{-10},
& \text{free-streaming region},\\[4pt]
1.3\times10^{-7},
& \text{degenerate core}.
\end{cases}
\label{eq:multibody}
\end{equation}
This should be distinguished from the infrared-sensitive vacuum
multibody problem discussed in the 1990s. The large vacuum
many-body contribution proposed in Ref.~\cite{Fischbach1996} arose
from extrapolating a perturbative expansion into a regime in which
that expansion ceases to be reliable. For the hard background modes
considered here, Eq.~\eqref{eq:multibody} instead shows that additional
matter insertions remain perturbative. The two problems are controlled
by different expansion parameters and by different infrared physics.

\section{Robustness of the mean-field treatment}
\label{sec:robustness}

The mean-field estimate relies on three ingredients: the plane-wave pair
potential adopted from Ref.~\cite{Ghosh2023}, a target density that is
smooth on the microscopic scales relevant to the kernel, and a neutrino
background that is quasistatic on the microscopic phase-evolution
timescales. We examine each in turn.

\subsection{Plane-wave benchmark}

The present analysis takes the plane-wave pair potential of
Ref.~\cite{Ghosh2023} as its starting point. As a normalization check,
we reproduce that work's reactor benchmark
$|V_{\rm bkg}|/|V_{\rm grav}|\simeq2\times10^{-13}$ to within a
factor of two. Within this plane-wave treatment, the PNS pairwise
enhancement is very large.

The magnitude of the directional enhancement is itself under
discussion: Ref.~\cite{Blas2023} argues that the finite wave-packet
width of the background neutrinos suppresses it relative to the
plane-wave estimate. We adopt the plane-wave result throughout, which
is conservative for the present purpose. Any additional wave-packet
suppression would further reduce the predicted interaction and
therefore strengthen the conclusion of negligible PNS effects.

The new element considered here is the bulk PNS Hartree sum. The pair
potential $V(\mathbf r)$ at fixed separation and the volume integral
$\int d^3r\,V(\mathbf r)$ are logically distinct quantities.
Reference~\cite{Ghosh2023} already notes that averaging over
macroscopic test bodies smears the oscillatory force. Here we evaluate
the local homogeneous Hartree interaction over pair orientations and
spin structures and determine its scaling with the size of the
approximately homogeneous region. It is this bulk quantity, rather
than the largest fixed-direction pair potential, that is relevant to a
mean-field EOS correction or a volume-force estimate.

This conclusion does not apply directly to laboratory measurements
between finite test masses, whose geometry differs from the homogeneous
PNS Hartree problem. Section~\ref{sec:lab} discusses that geometry
separately.

\subsection{Density inhomogeneity and short-range correlations}

The local Hartree treatment assumes that the neutron density varies
slowly over the microscopic scales that dominate the oscillatory
kernel. Macroscopic inhomogeneity is not expected to alter the result.
Convective cells, turbulent eddies, and accretion structures vary on
metres to kilometres, vastly larger than the relevant microscopic
scales. The kernel varies on scales set by $E_\nu^{-1}$ or
$\mu_\nu^{-1}$, whereas the background density and angular
distribution vary on macroscopic hydrodynamic scales. The PNS can
therefore be treated locally as a sequence of approximately homogeneous
regions, with gradient corrections suppressed by the ratio of
microscopic to macroscopic length scales.

Short-range nuclear correlations are relevant in the core because the
SI integral receives substantial weight from separations of a few
femtometres. Examining where Eq.~\eqref{eq:SIlimit} accumulates for the
degenerate benchmark, we find that the dominant contribution lies at
$r\simeq1$--$3\,\mathrm{fm}$, with about $30\%$ accumulated below
$1\,\mathrm{fm}$ and $99\%$ below $3\,\mathrm{fm}$. This is precisely
the range over which the neutron--neutron pair correlation function
$g_{nn}(r)$ can depart appreciably from unity.

At the pair-distribution level, the Hartree weighting is therefore
modified according to
$n_n^2\to n_n^2g_{nn}(r)$. Rather than adopt a single correlation
function, we scan the Jastrow-like family
\begin{equation}
g_{nn}(r)
=
\left[1-e^{-(r/r_c)^2}\right]^2
\left[
1+a_J e^{-((r-r_1)/w)^2}
\right],
\label{eq:jastrow}
\end{equation}
where $r_c$ sets the correlation-hole scale, while $a_J$, $r_1$, and
$w$ control the amplitude, position, and width of the first correlation
peak. We use the grid
\begin{align}
r_c&\in
\{0.4,0.5,0.6,0.7,0.8,0.9,1.0\}\ {\rm fm},
\nonumber\\
a_J&\in
\{0,0.15,0.30,0.50\},
\nonumber\\
r_1&\in
\{1.2,1.6,2.0\}\ {\rm fm},
\nonumber\\
w&\in
\{0.3,0.5,0.8\}\ {\rm fm},
\end{align}
chosen to span the correlation-hole radius and the position and width
of the first correlation peak typical of variational nuclear-matter
calculations~\cite{Pandharipande1979}. Across the
$7\times4\times3\times3=252$ parameter combinations, the SI Hartree
value is rescaled by
\begin{equation}
\frac{
\int d^3r\,g_{nn}(r)V
}{
\int d^3r\,V
}
\in
[0.62,\,1.22],
\end{equation}
with median $0.87$.

Thus short-range central correlations produce an $O(1)$ correction
across this representative family. The spread is relevant for a
precision calculation but immaterial to the qualitative conclusion,
since the uncorrelated benchmark is already suppressed by
$4.5\times10^{-15}$ relative to $P_{\rm deg}$.

Correlations also break the infinite-volume uncorrelated SD-PC
cancellation of Eq.~\eqref{eq:sdpczero}. With
$g_{nn}(r)\neq1$, the radial moment identity is no longer preserved:
\begin{equation}
\int dr\,
g_{nn}(r)
\frac{
\mathcal J_b+\tfrac13\mathcal J_c
}{r^3}
\neq0
\end{equation}
in general. Repeating the same 252-point parameter scan and normalizing
to the correspondingly correlated SI kernel gives
\begin{equation}
\mathcal R^{\rm SD-PC}_{\rm corr}
\equiv
\frac{
\displaystyle
\int_0^\infty
\frac{dr}{r^3}\,
g_{nn}(r)
\left[
\mathcal J_b(r)
+\tfrac13\mathcal J_c(r)
\right]
}{
\displaystyle
\int_0^\infty
\frac{dr}{r^3}\,
g_{nn}(r)\mathcal J_a(r)
}
\in[-0.31,\,0.02],
\label{eq:sdpccorr}
\end{equation}
with median $-0.14$.

A central Jastrow correlation function does not exhaust neutron-matter
correlations. As an independent benchmark, we therefore also consider
the free Fermi-gas exchange hole of identical unpolarized neutrons,
\begin{equation}
g_{nn}^{\rm FG}(r)
=
1-\frac12
\left[
\frac{3j_1(k_Fr)}{k_Fr}
\right]^2,
\end{equation}
where $j_1$ is the spherical Bessel function of order one. This
rescales the SI integral by $0.86$ and gives
\begin{equation}
\mathcal R^{\rm SD-PC}_{\rm corr}
=
-0.08,
\end{equation}
within the Jastrow band.

Equation~\eqref{eq:sdpccorr} is a ratio of kernel integrals; coupling
and spin factors are not included. In a factorized polarized mean
field, the physical SD-PC/SI ratio therefore carries the additional
factor
\begin{equation}
\frac{(g_A^n)^2P^2}{(g_V^n)^2},
\end{equation}
with $P^2\sim10^{-7}$ for the benchmark polarization. This
factorization is itself a mean-field statement, because genuine spin
correlations can make
$\langle\bsig_1\cdot\bsig_2\rangle$ differ from the product of the
one-body spin expectation values. The conclusion is insensitive to
this assumption: even if spin correlations removed the $P^2$
suppression, the resulting SD-PC contribution would remain at most
$O(1)$ relative to the already negligible SI interaction.

At $g_{nn}(r)=1$, the exact zero in Eq.~\eqref{eq:sdpczero} is the
infinite homogeneous result. For a finite homogeneous core, the
nonzero boundary remainder is given by Eq.~\eqref{eq:sdpcfinite} and is
already $O[(\mu_\nu R)^{-1}]\sim10^{-20}$ at the PNS benchmark.
Short-range correlations therefore produce a much larger correction
than this formal finite-size remainder, but only at the $O(1)$ kernel
level relative to an already negligible SI contribution.

In the free-streaming region, the exact SI form factor settles the
correlation question directly. From Eq.~\eqref{eq:master}, the
fraction of the saturated monochromatic-beam result accumulated inside
radius $R$ is
\begin{equation}
\mathcal F(E_\nu R)
=
1-\frac{\sin(2E_\nu R)}{2E_\nu R}.
\end{equation}
At $R=3\,\mathrm{fm}$ and $E_\nu=12\,\mathrm{MeV}$,
\begin{equation}
\mathcal F(E_\nu R)\simeq0.022.
\end{equation}
Only a few percent of the uncorrelated free-streaming integral
therefore resides inside the nuclear-correlation region, and
representative short-range correlation profiles modify the saturated
result only at the percent level.

\subsection{Time dependence of the background}

We treat $\Phi_0$, $E_\nu$, and $\khat$ as quasistatic over the
microscopic phase-evolution times $E_\nu^{-1}$ and $\mu_\nu^{-1}$.
For $E_\nu=12\,\mathrm{MeV}$,
\begin{equation}
E_\nu^{-1}
\simeq
5.5\times10^{-23}\,\mathrm{s},
\end{equation}
while in the core
\begin{equation}
\mu_\nu^{-1}
\simeq
3.3\times10^{-24}\,\mathrm{s}.
\end{equation}
These microscopic timescales are therefore approximately
$10^{19}$--$10^{20}$ times shorter than millisecond-scale variations
of the PNS background. The background is consequently adiabatic on
the phase-evolution times relevant to the neutrino-mediated kernel.

\section{Comparison with weak effects}
\label{sec:hierarchies}

Two additional scale comparisons reinforce the smallness of the
background-enhanced pair-exchange effect.

\emph{(i) Incoherent scattering dominates.}
For low-energy neutral-current elastic scattering on neutrons, the
leading cross section is
\begin{equation}
\sigma_{\nu n}
=
\frac{\GF^2\Enu^2}{\pi}
\left(c_V^2+3c_A^2\right)
=
3.5\times10^{-42}\,
\mathrm{cm^2},
\end{equation}
at $\Enu=12\,\mathrm{MeV}$, where
\begin{equation}
c_V=-\frac12,
\qquad
c_A=-\frac{g_A^{(N)}}{2}\simeq-0.635,
\qquad
g_A^{(N)}\simeq1.27.
\end{equation}
For the free-streaming benchmark, a characteristic momentum-transfer
force per neutron is therefore
\begin{equation}
F_{\rm scat}^{(1)}
\simeq
(cn_\nu)\,
\sigma_{\nu n}\,
\frac{\Enu}{c}
\simeq
3.1\times10^{-14}\,\mathrm{dyn}.
\end{equation}
By comparison, the background-enhanced pair interaction produces the
mean-field force scale
\begin{equation}
F_{2\nu}^{(1)}
\equiv
\frac{|U_{2\nu}|}{H}
\simeq
2.0\times10^{-32}\,\mathrm{dyn},
\end{equation}
for the same benchmark. Thus
\begin{equation}
\frac{F_{\rm scat}^{(1)}}{F_{2\nu}^{(1)}}
\simeq
1.6\times10^{18}.
\end{equation}
Charged-current absorption can provide a still larger momentum-transfer
rate for the relevant flavor components, but the neutral-current
estimate alone is sufficient to establish the hierarchy.

\emph{(ii) Lower-order refraction dominates.}
As a characteristic lower-order coherent weak scale, consider the
neutrino-background forward-scattering potential
\begin{equation}
V_{\nu\nu}^{\rm ref}
\equiv
\sqrt2\,\GF n_\nu
=
5.8\times10^{-5}\,\mathrm{eV}
\end{equation}
at the free-streaming benchmark density. This exceeds the magnitude of
the pair-exchange mean-field potential,
\begin{equation}
|U_{2\nu}|
=
1.8\times10^{-14}\,\mathrm{eV},
\end{equation}
by
\begin{equation}
\frac{V_{\nu\nu}^{\rm ref}}{|U_{2\nu}|}
\simeq
3\times10^9.
\end{equation}
The actual neutrino--neutrino refractive potential depends on the
flavor composition, the neutrino--antineutrino asymmetry, and, for an
anisotropic background, the relative propagation angles. The quantity
$\sqrt2\,\GF n_\nu$ is therefore quoted only as a characteristic weak
mean-field scale.

The perturbative ordering is nevertheless transparent. The
background-enhanced pair interaction is of order $\GF^2$, whereas
ordinary forward scattering is of order $\GF$. Their hierarchy contains
the additional small parameter
\begin{equation}
\frac{\GF n_n}{\Enu}\ll1
\end{equation}
in the free-streaming PNS benchmark, consistent with the expansion
parameter discussed in Eq.~\eqref{eq:multibody}.

\section{Pulsar kicks}
\label{sec:kicks}

Observed young pulsars have characteristic birth velocities of several
hundred $\mathrm{km\,s^{-1}}$, with a high-velocity tail extending
beyond $1000\,\mathrm{km\,s^{-1}}$~\cite{Hobbs2005}. We next estimate the largest pulsar recoil that could be generated by
the background-enhanced interaction. A direct force estimate and an
independent neutrino momentum-budget bound both give negligible effects.

For the direct estimate, let $\epsilon_F$ denote the dipole fraction of
the coherent force acting on a shell of mass $M_{\rm shell}$. Using the
one-nucleon force scale $F_{2\nu}^{(1)}$ from
Sec.~\ref{sec:hierarchies}, the dipolar force is
\begin{equation}
F_z
\simeq
\epsilon_F
\frac{M_{\rm shell}}{m_n}
F_{2\nu}^{(1)} .
\end{equation}
For
\begin{equation}
\epsilon_F=10^{-8},
\qquad
M_{\rm shell}=10^{-4}M_\odot,
\end{equation}
this gives
\begin{equation}
F_z
\simeq
3\times10^{13}\,\mathrm{dyn}.
\end{equation}
Applied over a characteristic neutrino-emission time
$\tau\simeq10\,\mathrm{s}$, the corresponding recoil velocity is
\begin{equation}
\Delta v
\simeq
\frac{F_z\tau}{M}.
\end{equation}
Even taking the limiting value $\epsilon_F=1$ and increasing
$M_{\rm shell}$ by a factor of $100$ gives only
$\Delta v=O(10^{-11})\,\mathrm{m\,s^{-1}}$ for a
$1.4M_\odot$ remnant.

An independent upper bound follows from momentum conservation. Taking
the total energy emitted in neutrinos during the cooling epoch to be
\begin{equation}
E_\nu^{\rm tot}
\simeq
3\times10^{53}\,\mathrm{erg},
\end{equation}
the total neutrino momentum scale is
\begin{equation}
p_\nu
=
\frac{E_\nu^{\rm tot}}{c}
\simeq
10^{43}\,
\mathrm{g\,cm\,s^{-1}}.
\end{equation}
Let $\epsilon_\nu$ denote the dipole anisotropy of the emitted neutrino
momentum. A mechanism that transfers momentum between the neutrino
radiation and the stellar matter cannot impart a net stellar momentum
larger than the corresponding anisotropic neutrino component,
\begin{equation}
\Delta p_{\rm NS}
\lesssim
\epsilon_\nu p_\nu .
\end{equation}
Hence a recoil speed $v_{\rm kick}$ requires
\begin{equation}
\epsilon_\nu
\gtrsim
\frac{M v_{\rm kick}}{p_\nu}.
\end{equation}
For a $1.4M_\odot$ remnant, reaching
$1000\,\mathrm{km\,s^{-1}}$ requires
\begin{equation}
\epsilon_\nu
\gtrsim
2.8\%,
\end{equation}
while $400\,\mathrm{km\,s^{-1}}$ requires
\begin{equation}
\epsilon_\nu
\gtrsim
1.1\%.
\end{equation}

The few-percent neutrino momentum anisotropy required for an
astrophysically relevant kick is therefore a kinematic lower bound,
independent of the microscopic mechanism that transfers the anisotropic
neutrino momentum to the star. The direct force estimate is far below
even this mechanism-independent bound.

\section{Beyond-Standard-Model reach}
\label{sec:bsm}

Consider first a simple rescaling of the same effective four-fermion
interaction, $\GF\to G_{\rm eff}$, with no change in the kernel
structure. All Hartree results then scale as $G_{\rm eff}^2$.
Requiring an order-unity effect relative to the benchmark pressure
scales gives
\begin{equation}
\frac{G_{\rm eff}}{\GF}
\gtrsim
1.5\times10^{7}(\text{core}), \
1.2\times10^{10}(\text{free-streaming region}).
\end{equation}
Defining
\begin{equation}
\varepsilon_{\rm NSI}
\equiv
\frac{G_{\rm eff}}{\GF},
\end{equation}
representative global analyses constrain many conventional contact
non-standard-interaction coefficients to $O(1)$ or below, with the
precise bounds depending on flavor and operator structure
~\cite{Coloma2023}. The effective couplings required for PNS relevance
are therefore many orders of magnitude larger than those allowed for
conventional contact NSI scenarios.

The non-accumulation result is not universal to all BSM interactions.
A background-induced model must be checked against three separate
conditions. First, the background enhancement should arise from an
on-shell contribution that produces an oscillatory position-space
kernel. Second, the relevant spectral support must lie in the
short-wavelength regime of Eq.~\eqref{eq:validity}. Third, the
Hartree-averaged kernel must provide sufficient angular or radial
suppression to remove any residual $R$-dependence;
Eq.~\eqref{eq:counterexample} shows explicitly that oscillation alone
is insufficient.

For the SM SI interaction, the third property follows from the angular
projection Eq.~\eqref{eq:avgidentity} and the exact form factor
Eq.~\eqref{eq:formfactor}. A BSM model may preserve the on-shell
enhancement and short-wavelength behavior while modifying the numerator
or angular structure. Such a model need not exhibit the same Hartree
saturation, and its bulk kernel must therefore be analyzed separately.

Macroscopic $R^2$ accumulation can instead arise from a sufficiently
light tree-level mediator. If $m_\phi$ denotes the mediator mass, the
condition
\begin{equation}
m_\phi R\ll1
\end{equation}
makes its Yukawa potential effectively Coulombic across the system,
so that the Hartree integral grows as $R^2$. This is a conventional
long-range fifth-force scenario rather than the background-induced
mechanism considered here.

Reference~\cite{XuYu2022} catalogues BSM modifications of the
\emph{vacuum} neutrino force when the effective contact interaction is
resolved by additional mediators. Background-enhanced versions of such
models must be checked independently against the three conditions above.

\section{Implications for laboratory searches}
\label{sec:lab}

The directional coherence condition
\begin{equation}
\theta^2
\lesssim
\frac{2}{\Enu r}
\end{equation}
places simultaneous requirements on the angular extent of the neutrino
source and on the geometry of the test masses. For a Galactic
supernova at $10\,\mathrm{kpc}$, a neutrinosphere with radius of order
tens of kilometres has an angular radius of order
$10^{-16}$--$10^{-17}\,\mathrm{rad}$. Its finite angular extent is
therefore negligible for laboratory separations. By contrast, the Sun
has an angular radius of order $10^{-2}\,\mathrm{rad}$, and for
MeV-scale solar neutrinos this angular spread exceeds the narrow
forward-coherence requirement by many orders of magnitude.

Let $r$ denote the characteristic longitudinal separation of a pair
and $d$ the transverse extent of the test masses. Since the associated
angular spread is of order $\theta\sim d/r$, the coherence condition
requires approximately
\begin{equation}
d
\lesssim
\sqrt{\frac{2r}{\Enu}}.
\end{equation}
For $\Enu=12\,\mathrm{MeV}$ this gives
\begin{equation}
d\lesssim18\,\mathrm{nm}
\qquad
(r=1\,\mathrm{cm}),
\end{equation}
\begin{equation}
d\lesssim180\,\mathrm{nm}
\qquad
(r=1\,\mathrm{m}),
\end{equation}
and
\begin{equation}
d\lesssim570\,\mathrm{nm}
\qquad
(r=10\,\mathrm{m}).
\end{equation}
These transverse dimensions are compatible with levitated-nanosphere
scales, although the predicted interaction remains extremely small.

Atomic and spectroscopic probes of the \emph{vacuum} neutrino force
have been developed in
Refs.~\cite{Dzuba2018,Ghosh2020,GhoshAllScales}. Their phenomenology
is distinct from that of the background-enhanced macroscopic
interaction considered here.

\section{Conclusions}
\label{sec:conclusions}

We have shown that the background-enhanced two-neutrino interaction
does not develop a macroscopically growing Hartree contribution under
the proto-neutron-star conditions considered here. The same on-shell
contribution that enhances the directional pair potential also produces
an oscillatory position-space kernel. For the Standard Model
interaction, the angular and radial structure of this kernel prevents
the large pairwise enhancement from accumulating over macroscopic
distances.

We note the general result obtained in the spin-independent
sector. Averaging over pair orientations projects an arbitrary
ultrarelativistic background onto its angular monopole. A directional
beam and an isotropic shell with the same monopole spectrum therefore
give the same homogeneous SI Hartree contribution. For each spectral
mode, the finite-$R$ result enters the short-wavelength regime when
$kR\gg1$ and approaches an $R$-independent limit because of the
specific angular structure of the SM kernel. The spin-dependent
sectors require a separate treatment, but likewise generate no growing
Hartree contribution under the PNS mean-field configurations considered
here. Short-range nuclear correlations can modify the core Hartree
kernel at the $O(1)$ level, but this is far too small to alter the
astrophysical conclusion.

The numerical difference is correspondingly large. In the
free-streaming region, the forward neutrino-mediated pair potential
exceeds the neutron--neutron gravitational pair potential by
$7.8\times10^{16}$, whereas the resulting one-body Hartree potential
is only $2.9\times10^{-22}$ of the stellar gravitational potential per
neutron. In the degenerate core, the magnitude of the SI interaction
energy density is only $4.5\times10^{-15}$ of the free-neutron-gas
degeneracy pressure. Lower-order weak processes are larger still. The
background-enhanced two-neutrino interaction is therefore dynamically
negligible for the PNS equation of state and cannot provide an
appreciable pulsar-kick contribution within the Standard Model
mechanism and benchmark conditions considered here.

This result should not be interpreted as a general statement about all
background-induced forces. Macroscopic accumulation depends not only
on the presence of an on-shell enhancement, but also on the wavelength
of the background modes and on the angular and radial structure of the
Hartree-averaged kernel. Oscillation alone is insufficient. For a
macroscopic medium, the relevant diagnostic is therefore the bulk
Hartree integral and its scaling with system size, rather than the
largest fixed-direction pair potential.

\section*{Data availability}

No external datasets were used. All numerical results were generated from the analytic expressions, benchmark parameters, and parameter grids specified in the text, from which they can be reproduced directly.

\begin{acknowledgments}
The authors acknowledge the use of generative-AI tools for language refinement to improve the readability of this manuscript. All content and conclusions presented herein remain the sole responsibility of the authors.
\end{acknowledgments}

\end{document}